%% file: main.tex
\documentclass[runningheads]{llncs}
\usepackage[T1]{fontenc}
\usepackage{graphicx}
\usepackage{booktabs}
\usepackage[misc]{ifsym}

\usepackage{multirow}
\usepackage{algorithm}
\usepackage[noend]{algpseudocode}
\usepackage{makecell}
\usepackage{amssymb}
\usepackage{tikz}
\usepackage[most]{tcolorbox}
\usepackage{xcolor}
\usetikzlibrary{positioning, shapes.symbols, calc}
\tikzset{puzzle/.style={fill=#1, inner sep=2pt, rounded corners=1pt, rotate=45, minimum size=0.3cm}}
\usepackage{hyperref}
\usepackage{subcaption}

\definecolor{indBlue}{HTML}{4089A9}
\definecolor{transGray}{HTML}{8F9A90}
\definecolor{stateFill}{HTML}{EFEFED}
\definecolor{stateStroke}{HTML}{A8A8A4}
\definecolor{headerFill}{HTML}{FAF9F7}
\definecolor{panelBorder}{HTML}{D4CFC8}
\definecolor{inkLight}{HTML}{6B6560}
\definecolor{innerBoxFill}{HTML}{E8F0E9}
\definecolor{errorRed}{HTML}{CC0000}

\usepackage{acronym}
\acrodef{pbe}[PBE]{Programming-by-example}
\acrodef{io}[I/O]{input-output}
\acrodef{dsl}[DSL]{Domain-Specific Language}
\acrodef{ai}[AI]{Artificial Intelligence}
\acrodef{agi}[AGI]{Artificial General Intelligence}
\acrodef{arc}[ARC-AGI]{Abstraction and Reasoning Challenge}
\acrodef{regesm}[REGISM]{Repeated Execution-Guided Invoking of Synthesis Model}
\acrodef{tiips}[TIIPS]{Transductively Informed Inductive Program Synthesis}
\acrodef{llm}[LLM]{Large Language Model}
\acrodef{arc}[ARC-AGI]{Abstraction and Reasoning}
\acrodef{loo}[LOO]{leave-one-out}

\begin{document}
\title{Beyond Either-Or Reasoning: Transduction and Induction as Cooperative Problem-Solving Paradigms}

\titlerunning{Transduction and Induction as Cooperative Problem-Solving Paradigms}
\author{Janis Zenkner{(\Letter)} \orcidID{0009-0007-5088-8692} \and Tobias Sesterhenn \orcidID{0009-0008-1995-2294} \and Christian Bartelt \orcidID{0000-0003-0426-6714}}
\institute{Clausthal University of Technology, Clausthal-Zellerfeld, Germany \email{janis.zenkner@tu-clausthal.de}}
\authorrunning{J. Zenkner et al.}

\tocauthor{Janis Zenkner, Tobias Sesterhenn, Christian Bartelt}
\toctitle{Beyond Either-Or Reasoning: Transduction and Induction as Cooperative Problem-Solving Paradigms}

\maketitle              

\begin{abstract}
Traditionally, in \ac{pbe} the goal is to synthesize a program from a small set of input-output examples.
Lately, \ac{pbe} has gained traction as a few-shot reasoning benchmark, relaxing the requirement to produce a program artifact altogether which allows transductive methods to directly predict the missing output sample.
Transduction and induction are complementary reasoning modes -- where induction derives general rules from examples, transduction leverages the examples directly to infer specific outputs without intermediate generalization.
Yet existing approaches either treat them as mutually exclusive or couple them in hybrid structures where one paradigm dictates a fixed trajectory for the other -- undermining the latter's reasoning potential and creating cascading errors. 
We move away from these hierarchical models and introduce cooperative transductive-inductive problem solving: by interleaving both reasoning modes and ensuring neither unconditionally dominates the other, we preserve the search autonomy and reasoning capacity of each paradigm.
We instantiate this concept in \acs{tiips}.
Across three \ac{pbe} domains, \acs{tiips} consistently outperforms state-of-the-art baselines and generates programs that more closely mirror ground-truth trajectories in both syntax and semantics, indicating a better match to the intended program behavior.
Our findings highlight cooperative reasoning as a promising new direction for harnessing the full power of inductive and transductive reasoning.\footnote{Code can be found at \href{https://github.com/jzenkner/CooperativeReasoning}{https://github.com/jzenkner/CooperativeReasoning}.}.

\end{abstract}

\acresetall

\input{chapters/01intro}
\input{chapters/03relatedwork}
\input{chapters/04concept}
\input{chapters/05methods}
\input{chapters/06evaluation}
\input{chapters/07results}
\input{chapters/08conclusion}
\bibliographystyle{splncs04}
\bibliography{main}

\end{document}

%% file: chapters/01intro.tex
\section{Introduction}
\label{sec:intro}
Program synthesis aims to automatically generate programs from high-level specifications~\cite{solar2008program}.
Within this domain, \ac{pbe} has emerged as a vital subfield, in which tasks are specified via \ac{io} examples rather than formal logical descriptions or natural language~\cite{devlin2017robustfill}. 
Formally, a \ac{pbe} task is defined by a set of input-output pairs $\mathcal{E} = \{(I_i, O_i)\}_{i=1}^n$ and the goal of a solver is to find a latent function $p$ such that $p(I_i) = O_i$ $\forall (I_i, O_i) \in \mathcal{E}$.
To make this concrete, consider a string-manipulation task: given the examples  \texttt{("John Smith", "J. Smith")} and \texttt{("Ada Lovelace", "A. Lovelace")}, the solver must infer the underlying transformation and predict the missing output -- without ever being told what the transformation is. 
We provide a formal treatment in the supplementary material.\footnote{Appendix can be found \href{https://jzenkner.github.io/CooperativeReasoning/}{https://jzenkner.github.io/CooperativeReasoning/}.}
While this principle improves the accessibility of software development~\cite{gulwani2011automating} -- end users can specify intent by demonstration rather than by writing code -- it introduces a core challenge: \ac{io} specifications are rarely complete, so many programs are consistent with the observed examples, and the solver must recover the intended latent function from an inherently underspecified signal~\cite{gulwani2017program}.

Two fundamentally different strategies have emerged for doing so.
The traditional \emph{inductive} approach treats \ac{pbe} as a program synthesis problem: search over a space of programs until one is found that is consistent with the \ac{io} examples.
This yields a reusable, explicit program that can be applied to any future input and inspected for correctness. 
More recently, \emph{transductive} approaches have reframed \ac{pbe} as a few-shot learning problem: rather than constructing an explicit program, the \ac{io} examples serve as context for a direct input-to-output mapping, and the missing output is predicted without ever producing a program artifact~\cite{chollet2024arc,cole2024machine}. 
Both strategies are viable, and each has been shown to work well in practice.

\begin{figure}[t]
    \centering
    \subcaptionbox{Subtask favoring induction\label{fig:induction}}{
        \includegraphics[width=0.47\linewidth]{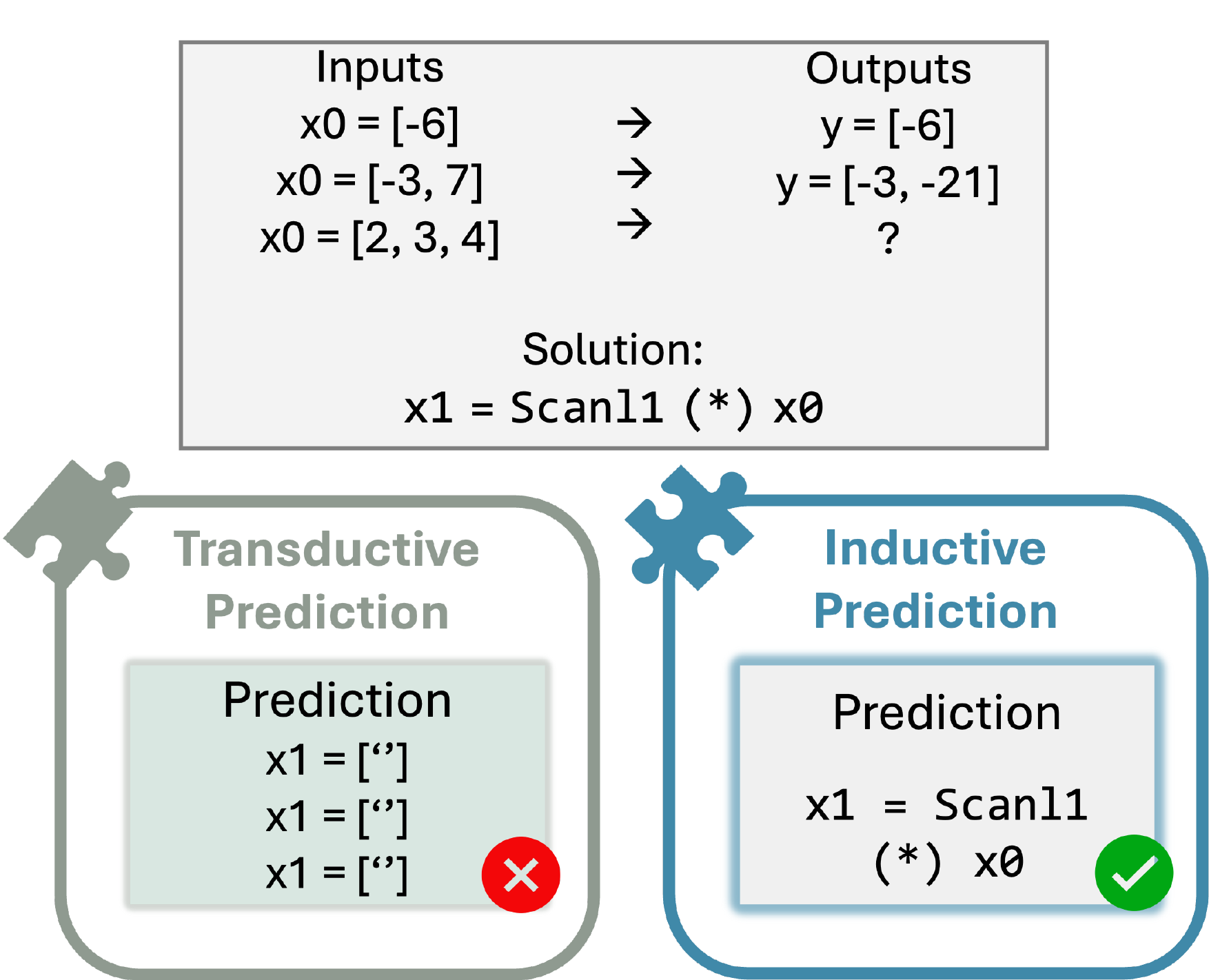}
    }
    \subcaptionbox{Subtask favoring transduction\label{fig:transduction}}{
        \includegraphics[width=0.47\linewidth]{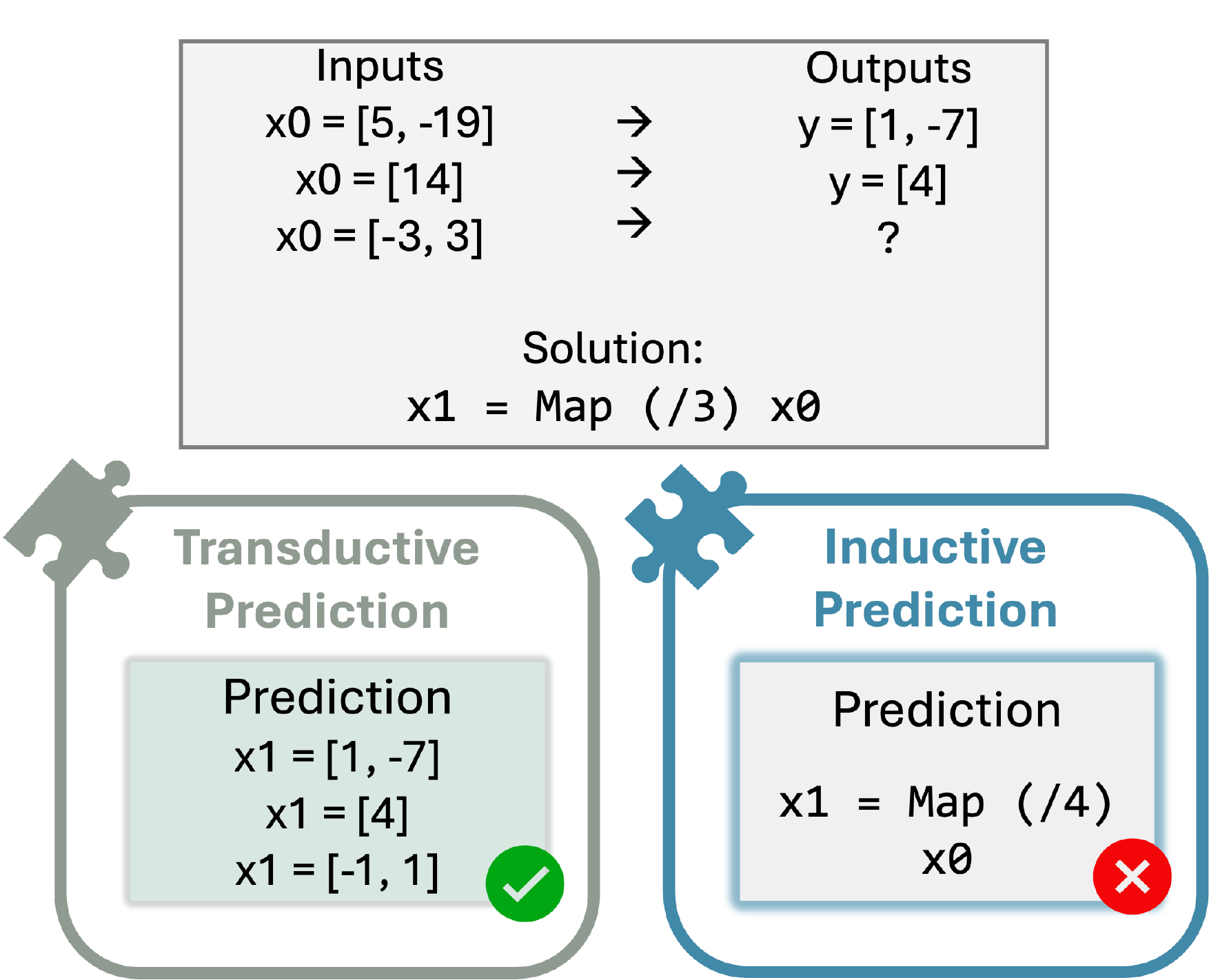}
    }
    \caption{Real model predictions on two DeepCoder list-manipulation subtasks~\cite{balog2016deepcoder}, illustrating complementarity between induction and transduction. Each panel shows a single-step \ac{pbe} task: given the \ac{io} examples, predict the missing output. The transductive model predicts (intermediate) outputs directly; the inductive model generates an explicit (sub-)program. (a) The transductive model produces malformed predictions, while the inductive model correctly synthesizes \texttt{Scanl1 (*) x0} to compute the cumulative product of the list. (b) The transductive model correctly predicts the output, while the inductive model synthesizes an incorrect subprogram \texttt{Map (/4) x0} dividing each element by four instead of \texttt{Map (/3) x0}. Note, floats are casted to integers. These are real outputs from both models -- not constructed for illustration.}
    \label{fig:complementarity}
\end{figure}
While both strategies succeed, they fail on systematically different tasks~\cite{zenkner2025shedding}. 
As shown in~\cite{li2024combining}, transduction excels when the underlying rule is difficult to formalize, pattern-matching fluidly over examples in cases where explicit program construction stalls.
Induction is stronger where the transformation requires precise compositional reasoning, maintaining logical consistency and producing verifiable artifacts.
Fig.~\ref{fig:complementarity} makes this concrete: it shows real model predictions from both paradigms taken from our evaluation on two single-step list manipulation tasks. 
In Fig.~\ref{fig:induction}, the transductive model produces malformed predictions while the inductive model finds the correct program; in Fig.~\ref{fig:transduction}, the inductive model synthesizes an incorrect program while the transductive model correctly predicts the output.
This complementarity naturally raises the question of how both paradigms can be integrated effectively.

Existing work has sought to bridge this gap, but the resulting approaches remain unsatisfactory.
The reason becomes clear once the structure of \ac{pbe} tasks is taken into account.
Most \ac{pbe} tasks are inherently compositional: the latent function underlying a task can typically be decomposed into a sequence of simpler sub-functions, each transforming the input one step closer to the output. 
Crucially, this compositionality is a property of the \emph{task} itself -- independent of whether an explicit program is ever constructed or whether a transductive approach is used. 
It induces a natural \emph{solution trajectory}: a chain of intermediate states $s_0 \to s_1 \to \cdots \to s_n$, where $s_0$ is the input, $s_n$ is the target output, and each $s_i$ is produced by one sub-function. 
Under induction, each transition $s_i \to s_{i+1}$ corresponds to synthesizing a subprogram; under transduction, it corresponds to directly predicting the next intermediate state from the examples. 
Returning to Fig.~\ref{fig:complementarity}: the two tasks shown are not complete multi-step tasks but individual steps $s_i \to s_{i+1}$ drawn from longer solution trajectories -- what the figure illustrates is subtask-level behavior.
This upgrades the complementarity observation: induction and transduction do not merely fail on different tasks, they disagree on individual steps within the same solution trajectory.

\begin{figure}[t]
\centering
\begin{tikzpicture}[
  state/.style={
    circle, minimum size=0.7cm, inner sep=0pt,
    fill=stateFill, draw=stateStroke, line width=0.6pt
  },
  statedash/.style={
    circle, minimum size=0.7cm, inner sep=0pt,
    fill=stateFill, draw=stateStroke, line width=0.6pt, dashed
  },
  indarr/.style={->, >=stealth, line width=0.9pt, color=indBlue},
  transarr/.style={->, >=stealth, line width=0.9pt, color=transGray},
  nodelbl/.style={font=\small, text=inkLight},
  arrowlbl/.style={font=\small},
]


\def\bL{0}          
\def\bR{12}       
\def\hH{0.5}       
\def\bH{1.25}        
\def\rH{1.75}       
\def\rGap{0.1}     
\def\rS{1.8}       

\def\xA{1.25}
\def\xB{3.65}
\def\xC{6.05}
\def\xD{8.45}
\def\xE{10.85}

\def\tA{0}
\pgfmathsetmacro\tB{-\rS}
\pgfmathsetmacro\tC{-2*\rS}
\pgfmathsetmacro\tD{-3*\rS}

\pgfmathsetmacro\nA{\tA  - \hH - \bH/2}
\pgfmathsetmacro\nB{\tB  - \hH - \bH/2}
\pgfmathsetmacro\nC{\tC  - \hH - \bH/2}
\pgfmathsetmacro\nD{\tD  - \hH - \bH/2}

\newcommand{\rowbox}[3]{%
  \fill[white] (\bL,#1) rectangle (\bR, #1-\rH);
  \draw[draw=panelBorder, line width=0.5pt]
    (\bL,#1) rectangle (\bR, #1-\rH);
  \fill[fill=headerFill]
    (\bL, #1) rectangle (\bR, #1-\hH);
  \draw[draw=panelBorder, line width=0.4pt]
    (\bL, #1-\hH) -- (\bR, #1-\hH);
  \node[anchor=west, text=black, font=\small, inner sep=0pt]
    at (\bL+0.25, #1-\hH/2)
    {\textbf{#2} {-- #3}};
}

\rowbox{\tA}%
  {(a)~\textsc{Inductive}}%
  {inductive model holds full trajectory autonomy}

\rowbox{\tB}%
  {(b)~\textsc{Transductive}}%
  {transductive model holds full trajectory autonomy}

\rowbox{\tC}%
  {(c)~\textsc{Hybrid}}%
  {states fixed transductively; induction cannot deviate}

\rowbox{\tD}%
  {(d)~\textsc{Cooperative}}%
  {all paradigms contribute; neither  controls the trajectory}

\node[state] (a0) at (\xA,\nA) {};
\node[state] (a1) at (\xB,\nA) {};
\node[state] (a2) at (\xC,\nA) {};
\node[state] (a3) at (\xD,\nA) {};
\node[state] (a4) at (\xE,\nA) {};
\draw[indarr] (a0)--node[above,arrowlbl,color=indBlue]{\texttt{prog$_1$}}(a1);
\draw[indarr] (a1)--node[above,arrowlbl,color=indBlue]{\texttt{prog$_2$}}(a2);
\draw[indarr] (a2)--node[above,arrowlbl,color=indBlue]{\texttt{prog$_3$}}(a3);
\draw[indarr] (a3)--node[above,arrowlbl,color=indBlue]{\texttt{prog$_4$}}(a4);
\foreach \x/\s in {\xA/$s_0$,\xB/$s_1$,\xC/$s_2$,\xD/$s_3$,\xE/$s_4$}
  \node[nodelbl] at (\x,\nA-0.50) {\s};

\node[state]     (b0) at (\xA,\nB) {};
\node[statedash] (b1) at (\xB,\nB) {};
\node[statedash] (b2) at (\xC,\nB) {};
\node[statedash] (b3) at (\xD,\nB) {};
\node[statedash] (b4) at (\xE,\nB) {};
\draw[transarr] (b0)--(b1);
\draw[transarr] (b1)--(b2);
\draw[transarr] (b2)--(b3);
\draw[transarr] (b3)--(b4);
\foreach \x/\s in {\xA/$s_0$,\xB/$s_1$,\xC/$s_2$,\xD/$s_3$,\xE/$s_4$}
  \node[nodelbl] at (\x,\nB-0.50) {\s};

\node[state]     (c0) at (\xA,\nC) {};
\node[statedash] (c1) at (\xB,\nC) {};
\node[statedash] (c2) at (\xC,\nC) {};
\node[statedash] (c3) at (\xD,\nC) {};
\node[statedash] (c4) at (\xE,\nC) {};
\draw[transarr] (c0)--node[above,arrowlbl,color=transGray]{\texttt{prog$_1$}}(c1);
\draw[transarr] (c1)--node[above,arrowlbl,color=transGray]{\texttt{prog$_2$}}(c2);
\draw[transarr] (c2)--node[above,arrowlbl,color=transGray]{\texttt{prog$_3$}}(c3);
\draw[transarr] (c3)--node[above,arrowlbl,color=transGray]{\texttt{prog$_4$}}(c4);
\foreach \x/\s in {\xA/$s_0$,\xB/$s_1$,\xC/$s_2$,\xD/$s_3$,\xE/$s_4$}
  \node[nodelbl] at (\x,\nC-0.50) {\s};

\node[state]     (d0) at (\xA,\nD) {};
\node[state]     (d1) at (\xB,\nD) {};
\node[statedash] (d2) at (\xC,\nD) {};
\node[state]     (d3) at (\xD,\nD) {};
\node[statedash] (d4) at (\xE,\nD) {};
\draw[indarr]   (d0)--node[above,arrowlbl,color=indBlue]{\texttt{prog$_1$}}(d1);
\draw[transarr] (d1)--(d2);
\draw[indarr]   (d2)--node[above,arrowlbl,color=indBlue]{\texttt{prog$_2$}}(d3);
\draw[transarr] (d3)--(d4);
\foreach \x/\s in {\xA/$s_0$,\xB/$s_1$,\xC/$s_2$,\xD/$s_3$,\xE/$s_4$}
  \node[nodelbl] at (\x,\nD-0.50) {\s};


\pgfmathsetmacro\legYone{\tD - \rH - \rGap - 0.42}
\pgfmathsetmacro\legYtwo{\tD - \rH - \rGap - 0.92}
\def\mid{6.75}
\def\lLeft{1.25}    
\def\lRight{6.25}   

\node[state, minimum size=0.30cm] at (\lLeft, \legYone) {};
\node[anchor=west, font=\scriptsize, text=inkLight]
  at (\lLeft+0.27, \legYone) {State (from program execution)};

\node[statedash, minimum size=0.30cm] at (\lRight, \legYone) {};
\node[anchor=west, font=\scriptsize, text=inkLight]
  at (\lRight+0.27, \legYone) {State (transductively determined)};

\draw[indarr] (\lLeft-0.32, \legYtwo) -- ++(0.60, 0);
\node[anchor=west, font=\scriptsize, text=inkLight]
  at (\lLeft+0.35, \legYtwo) {induction};

\draw[transarr] (\lRight-0.32, \legYtwo) -- ++(0.60, 0);
\node[anchor=west, font=\scriptsize, text=inkLight]
  at (\lRight+0.35, \legYtwo) {transduction\,/\,transductively constrained};

\end{tikzpicture}

\caption{
  Illustration of solution trajectories under four paradigms.
  Circles denote intermediate states~$s_i$; arrows denote transitions.
  (a)~Inductive: the inductive synthesizer generates all programs without constraint.
  (b)~Transductive: the next state is predicted directly at each step;
  no programs are generated.
  (c)~Hybrid: target states are pre-determined by the transductive
  model (dashed) before inductive synthesis begins; the inductive model then generates
  programs constrained to reach those states and cannot deviate
  from the imposed trajectory.
  (d)~Cooperative: free inductive steps and transductive/transductively-guided steps interleave.
  Unlike the hybrid case, neither paradigm unconditionally controls the trajectory.%
}
\label{fig:trajectories}
\end{figure}
The trajectory view in Fig.~\ref{fig:trajectories} makes the failure of existing combination strategies precise.
Systems either follow a purely inductive or purely transductive trajectory, or combine them only superficially.
Ensemble methods select one paradigm's output over the other at the task level -- if either fails on any step, the whole solution fails, and no subtask-level recovery is possible.
Current hybrid approaches, such as decomposition-based synthesis~\cite{shi2023exedec}, go further by using transductive predictions to pre-determine the full solution trajectory: the inductive model then searches locally within each predicted subtask, constrained to reach the transductively fixed next state $s_{i+1}$.
This superficially combines both paradigms, but in practice places induction entirely under the control of transduction: The inductive model retains only local search autonomy within each predicted subtask, while the overall trajectory is unconditionally fixed by the transductive model.
When the transductive decomposition is wrong at any step $s_i$, the error propagates along the remaining trajectory and the inductive model cannot deviate to find a globally consistent solution, even if one exists. 

Rather than subordinating one paradigm to the other, we propose allowing both to \emph{cooperate}: at each step along the solution trajectory, the paradigm better suited to the current subtask takes the lead -- but neither locks in a trajectory the other must follow. 
This enables mid-trajectory recovery in both directions: if the inductive synthesizer stalls at some step $s_i \to s_{i+1}$, transduction can redirect the search; if a transductive prediction is noisy, induction retains the autonomy to find a globally consistent solution rather than being forced to follow a flawed blueprint. 
To the best of our knowledge, we are the first to formalize this cooperative interaction between transduction and induction for program synthesis. Our contributions are:
\begin{enumerate}
\item \textbf{Cooperative Framework:} We define a general framework in which inductive and transductive reasoning actively interleave along the solution trajectory, ensuring both paradigms contribute according to local competence rather than one merely sketching a fixed trajectory for the other to follow.
\item \textbf{\acs{tiips}:} We introduce \ac{tiips} as a proof-of-concept instantiation of this framework. Rather than pre-determining a fixed trajectory, \ac{tiips} uses transduction as a \emph{search horizon reset} -- a targeted intervention that bypasses syntactic bottlenecks and redirects search at steps where induction stalls, without irrevocably constraining the inductive model's subsequent exploration.
\end{enumerate}

%% file: chapters/03relatedwork.tex
\section{Related Work}
\label{sec:relwork}

\paragraph{Inductive synthesis}
Traditional \ac{pbe} approaches search for programs consistent with the provided examples~\cite{solar2008program,gulwani2011automating,alur2013syntax,feser2015synthesizing}, using \acp{dsl} to constrain the search space. 
These restricted programming languages define the set of available operations and their compositions, trading expressive power for tractability and a bias toward generalizable solutions.
While interpretable and reusable, these methods suffer from combinatorial explosion as \ac{dsl} size or target program length grows.
To mitigate this, a prominent line of work uses neural models to guide search more efficiently~\cite{balog2016deepcoder,yin2017syntactic,lee2018accelerating}, including top-down and bottom-up synthesis over partial programs~\cite{nye2019learning,ellis2021dreamcoder,ellis2019write,shi2023lambdabeam} and execution-guided approaches that condition search on intermediate states~\cite{murali2017neural,chen2020compositional,hong2021latent,klinger2023compositional,prasad2023adapt,zhang2023planning,witt2023divide,demirtacs2025generating}. 
More recently, \acp{llm} have been applied as inductive reasoners, generating abstract hypotheses from examples and translating them into executable programs~\cite{qiu2023phenomenal,li2024programming,piriyakulkij2024doing,olausson2023self,madaan2023self}.
Across all these variants, the solution trajectory is determined entirely by the inductive paradigm -- the transductive component is absent entirely.

\paragraph{Transductive Solvers}
Framing \ac{pbe} as a pure few-shot reasoning task has motivated a parallel line of work that bypasses program generation entirely, directly predicting the missing test output from the provided examples~\cite{cole2024machine}.
Approaches include training task-specialized models~\cite{jolicoeur2025less}, direct output generation via high-capacity \acp{llm}~\cite{wang2023hypothesis,cole2024machine}, and test-time training in which the \ac{llm} adapts to the task-specific transformation logic before predicting~\cite{akyurek2024surprising}. 
These methods often achieve strong performance on reasoning benchmarks, but produce no interpretable or reusable artifact, and lose logical consistency on long, multi-step transformations. 
Symmetrically to inductive approaches, the solution trajectory here is determined entirely by transductive reasoning.

\paragraph{Hybrid Approaches}
Recognizing the complementarity of both paradigms, recent work has attempted to combine them: though none achieves true cooperation. 
We identify three distinct strategies, each of which still reduces to a trajectory controlled by a single paradigm.
The first strategy uses induction purely as a data augmentation tool to improve a downstream transductive solver via test-time training~\cite{franzen2024llm} or employs transduction as a feature extractor for subsequent solvers~\cite{mao2019neuro,mao2025neuro}.
The first paradigm here is a preprocessing step, not an active participant in problem-solving; the solution trajectory is determined by a single paradigm.
The second strategy, exemplified by BARC~\cite{li2024combining}, runs a transductive and an inductive approach in parallel, selecting whichever succeeds. 
Yet this constitutes an ensemble rather than a true collaboration: the two paradigms never interact, and neither can compensate for the other's failure.
The third and most sophisticated strategy, exemplified by ExeDec~\cite{shi2023exedec}, uses transductive predictions to iteratively decompose the full task into subtasks: the transductive model determines the intermediate target states along the trajectory, and the inductive model is then responsible for finding a program that transitions from the current state to each transductively prescribed next state. 
While the inductive model retains local reasoning autonomy within each subtask, it has no influence over the trajectory itself -- that is fixed entirely by the transductive model. 

%% file: chapters/04concept.tex
\section{Cooperative Transductive-Inductive Problem Solving}
\label{sec:concept}
The related work reveals a consistent structural pattern: every existing approach, whether purely inductive, purely transductive, or hybrid, relinquishes trajectory control to a single paradigm. 
The cooperative framework proposed here breaks with this pattern entirely.
Rather than appointing one paradigm as the "manager" and the other as its "worker", we treat both as active, co-equal agents whose contributions interleave dynamically throughout the solution process.

We formalize problem-solving as navigation through a state space $\mathcal{S}$ toward a goal state $s_\text{goal}$ consistent with $\mathcal{E}$. A solution trajectory is a sequence of states $(s_0, s_1, \dots, s_k)$ where each transition represents an act of problem-solving agency. 
We define the two transition types available at any state $s_t$
\[\text{inductive transitions } E_I = s_t \xrightarrow{p} s_{t+1}\] where $p \in \mathcal{P}$ is a symbolic program fragment. 
These transitions are interpretable and verifiable, but subject to combinatorial explosion as program length or \ac{dsl} size grows and
\[\text{transductive transitions } E_T = s_t \xrightarrow{M} s_{t+1}\] where $M$ is a neural model prediction. 
These transitions can leap across large regions of the state space without an explicit symbolic trace, but lack the logical consistency guarantees of rule-based search.
When a transductive transition precedes an inductive one, the predicted state $s_{t+1}$ simply becomes the new search root; the transductive step can be understood as a special \ac{dsl} operation whose ``execution'' is the neural model's prediction.
At any state $s_t$, a solver has access to the transition function: \[\Delta(s_t) = E_I(s_t) \cup E_T(s_t)\] and a trajectory is considered solved whenever the generated outputs -- whether produced inductively or transductively -- match the specifications in $\mathcal{E}$.

An approach is \emph{cooperative} if and only if it satisfies the following three criteria, each of which distinguishes it from existing hybrid and ensemble approaches:
\begin{enumerate}
    \item \textbf{Dual Agency}: Both $E_I$ and $E_T$ function as active problem-solvers. Neither paradigm exists solely to constrain, sketch, or translate the output of the other. This rules out approaches where induction serves only as a preprocessing step for transduction.
    \item \textbf{Interleaved Granularity}: The system permits switching between $E_I$ and $E_T$ at the level of individual reasoning steps, not only at task initialization or termination. This rules out ensemble approaches, where paradigms run in parallel and never interact within a single solution trajectory.
    \item \textbf{Search Autonomy Preservation}: After a transition, the state $s_{t+1}$ becomes a new search root from which either solver explores the \emph{full} solution/program space, unconstrained by further consistency requirements. This is the critical distinction from existing hybrid approaches: rather than locking induction into a corridor defined by transductive predictions, each transductive transition acts as a search horizon reset -- bypassing a syntactic bottleneck without irrevocably pruning the paths that follow.
\end{enumerate}
Any system satisfying the three criteria above qualifies as cooperative, regardless of how switching between $E_I$ and $E_T$ is scheduled or which models instantiate them.

%% file: chapters/05methods.tex
\section{\acl{tiips}}
\label{sec:method}
We instantiate this cooperative framework in \ac{tiips} while keeping the terminal requirement of producing an explicit symbolic program.
In this way, \ac{tiips} maintains interpretability.

We formulate synthesis as navigation through a state space consistent with Sec.~\ref{sec:concept}. 
At step $t$, the solver state \[s_t = \{(I_{t,i}, O_{t,i})\}_{i=1}^n\] represents the remaining transformation to be solved. 
The initial state $s_0$ corresponds to the original \ac{pbe} task, and the goal state $s_\text{goal}$ is reached when produced outputs match the targets for all examples.
A solution is a composition of \ac{dsl} operations \[P = \pi_1 \circ \dots \circ \pi_k\] where each $\pi_t$ transforms $s_t$ into $s_{t+1}$.
The key to unifying both paradigms is the subtask
\[\sigma_t = \{(I_{t,i}, T_{t,i})\} =  \begin{cases}
  \{(I_{t,i}, O_i)\}    & \text{if inductive mode} \\
  \{(I_{t,i}, v_{t,i})\}   & \text{if transductive mode}
\end{cases}
\] 
which specifies the local objective at each step. 
In \emph{inductive mode}, the subtask targets the final outputs directly: allowing the inductive synthesizer to reason without constraints.
In \emph{transductive mode}, the targets are replaced by transductively predicted intermediate values $v_{t}$: forcing a transductively determined trajectory on the synthesizer.
This shared subtask interface allows both transition types $E_I$ and $E_T$ to operate over the same state space and the same synthesizer model, without any architectural changes.

\ac{tiips} consists of two components. 
The \emph{inductive synthesizer} $f_{\mathrm{syn}}$ takes the current state and active subtask and predicts the next \ac{dsl} operation, which is executed to yield $s_{t+1}$.
The \emph{transductive guide} $f_{\mathrm{gui}}$ takes the current state and directly predicts the next intermediate values across all examples. 

\begin{algorithm}[!b]
\caption{\ac{tiips} algorithm using the Incremental Intervention Schedule.}
\label{algo:tiips}
\begin{algorithmic}[1] 
    \Require Transductive intervention limit $J$, Maximum program length $K$
    
    \For{$j = 0$ to $J$}
        \State $P \gets []$
        \State $s \gets \{(I_i, O_i)\}_{i=1}^n$
        
        \For{$t = 1$ to $K$}
            \If{$t \leq j$} \algorithmiccomment{Guided Mode: Intervention is active}
                \State $\{v_i\} \gets f_{gui}(s)$ 
                \State $\sigma \gets \{(I_{t,i}, v_{t,i})\}_{i=1}^n$ 
            \Else \algorithmiccomment{Free Mode: Inductive agency is prioritized}
                \State $\sigma \gets \{(I_{t,i}, O_i)\}_{i=1}^n$
            \EndIf
            
            \State $\pi_t \gets f_{syn}(s, \sigma)$ \algorithmiccomment{Predict the next subprogram}
            \State $\{\hat{O}_i\}_{i=1}^n \gets \text{Execute}(\pi_t, \{I_{t,i}\}_{i=1}^n)$
            
            \If{$\{\hat{O}_i\}_{i=1}^n = \{O_i\}_{i=1}^n$} \algorithmiccomment{Check if global task is solved}
                \State \Return $P + [\pi_t]$ 
            \EndIf
            
            \State $P \gets P + [\pi_t]$ 
            \State $s \gets \text{Update}(\{I_{t, i}, O_{t, i}\}_{i=1}^n, \{\hat{O}_i\}_{i=1}^n)$ \algorithmiccomment{Update state to represent what is left to be done (domain-specific)}
        \EndFor
    \EndFor
    \State \Return \text{Failure} \algorithmiccomment{No program found within budget}
\end{algorithmic}
\end{algorithm}
\ac{tiips} controls the interaction between paradigms through an \emph{incremental intervention schedule}, shown in Alg.~\ref{algo:tiips}. 
The outer loop iterates over transductive intervention levels $j = 0, \dots, J$.
At iteration $j$, the first $j$ steps of the trajectory are guided: $f_{\mathrm{gui}}$ predicts the next intermediate values and the subtask is set accordingly.
All subsequent steps operate in inductive mode, where $f_{\mathrm{syn}}$ reasons without constraints toward the final outputs. 
Each iteration re-initializes from $s_0$, so no guided prediction from a previous iteration influences the current one.
This yields three regimes: $j=0$ corresponds to purely inductive trajectories; $j=J$ to fully transductive trajectories; and $0 < j < J$ to cooperative regime, where transductive and inductive steps coexist within the same trajectory. 

\begin{figure}[t]
\centering
\begin{tikzpicture}[
  state/.style={
    circle, minimum size=0.7cm, inner sep=0pt,
    fill=stateFill, draw=stateStroke, line width=0.6pt
  },
  statedash/.style={
    circle, minimum size=0.7cm, inner sep=0pt,
    fill=stateFill, draw=stateStroke, line width=0.6pt, dashed
  },
  indarr/.style={->, >=stealth, line width=0.9pt, color=indBlue},
  transarr/.style={->, >=stealth, line width=0.9pt, color=transGray},
  nodelbl/.style={font=\small, text=inkLight},
  arrowlbl/.style={font=\small},
]


\def\bL{0}          
\def\bR{12}       
\def\hH{0.5}       
\def\bH{1.25}        
\def\rH{1.75}       
\def\rGap{0.1}     
\def\rS{0.1}       

\def\xA{1.25}
\def\xB{3.65}
\def\xC{6.05}
\def\xD{8.45}
\def\xE{10.85}

\def\tA{0}
\pgfmathsetmacro\tD{-1*\rS}

\pgfmathsetmacro\nA{\tA  - \hH - \bH/2}
\pgfmathsetmacro\nD{\tD  - \hH - \bH/2}

\newcommand{\rowbox}[3]{%
  \fill[white] (\bL,#1) rectangle (\bR, #1-\rH);
  \draw[draw=panelBorder, line width=0.5pt]
    (\bL,#1) rectangle (\bR, #1-\rH);
  \fill[fill=headerFill]
    (\bL, #1) rectangle (\bR, #1-\hH);
  \draw[draw=panelBorder, line width=0.4pt]
    (\bL, #1-\hH) -- (\bR, #1-\hH);
  \node[anchor=west, text=black, font=\small, inner sep=0pt]
    at (\bL+0.25, #1-\hH/2)
    {\textbf{#2} {-- #3}};
}

\rowbox{\tA}%
  {~\textsc{\ac{tiips}}}%
  {both reasoning paradigms determine the trajectory}

\node[state] (a0) at (\xA,\nA) {};
\node[statedash] (a1) at (\xB,\nA) {};
\node[statedash] (a2) at (\xC,\nA) {};
\node[state] (a3) at (\xD,\nA) {};
\node[state] (a4) at (\xE,\nA) {};
\draw[transarr] (a0)--node[above,arrowlbl,color=transGray]{\texttt{prog$_1$}}(a1);
\draw[transarr] (a1)--node[above,arrowlbl,color=transGray]{\texttt{prog$_2$}}(a2);
\draw[indarr] (a2)--node[above,arrowlbl,color=indBlue]{\texttt{prog$_3$}}(a3);
\draw[indarr] (a3)--node[above,arrowlbl,color=indBlue]{\texttt{prog$_4$}}(a4);
\foreach \x/\s in {\xA/$s_0$,\xB/$s_1$,\xC/$s_2$,\xD/$s_3$,\xE/$s_4$}
  \node[nodelbl] at (\x,\nA-0.50) {\s};


\pgfmathsetmacro\legYone{\tD - \rH - \rGap - 0.42}
\pgfmathsetmacro\legYtwo{\tD - \rH - \rGap - 0.92}
\def\mid{6.75}
\def\lLeft{1.25}    
\def\lRight{6.25}   

\node[state, minimum size=0.30cm] at (\lLeft, \legYone) {};
\node[anchor=west, font=\scriptsize, text=inkLight]
  at (\lLeft+0.27, \legYone) {State (from program execution)};

\node[statedash, minimum size=0.30cm] at (\lRight, \legYone) {};
\node[anchor=west, font=\scriptsize, text=inkLight]
  at (\lRight+0.27, \legYone) {State (transductively determined)};

\draw[indarr] (\lLeft-0.32, \legYtwo) -- ++(0.60, 0);
\node[anchor=west, font=\scriptsize, text=inkLight]
  at (\lLeft+0.35, \legYtwo) {induction};

\draw[transarr] (\lRight-0.32, \legYtwo) -- ++(0.60, 0);
\node[anchor=west, font=\scriptsize, text=inkLight]
  at (\lRight+0.35, \legYtwo) {transductively constrained};

\end{tikzpicture}

\caption{
  Illustration of \ac{tiips} trajectories.
  Circles denote intermediate states~$s_i$; arrows denote transitions.
  Transduction actively contributes to the solution by guiding the inductive solver's trajectory on the first $j$ steps while induction can reason freely on the remaining $K - j$ steps.
}
\label{fig:tiips_traj}
\end{figure}
The structural difference from hybrid approaches such as ExeDec~\cite{shi2023exedec} and ensemble approaches such as BARC~\cite{li2024combining} is sharp: in both cases the solution trajectory is ultimately controlled by a single paradigm, and a wrong prediction by that paradigm cannot be recovered.
In \ac{tiips}, by contrast, each transductive prediction influences only the immediate next subtask.
Once $s_{t+1}$ is reached, the inductive synthesizer explores the full program space from that state with no further prescription, making transductive transitions \emph{search horizon resets} rather than trajectory blueprints.

\ac{tiips} satisfies all cooperation criteria of Sec.~\ref{sec:concept}. 
Dual Agency is preserved because $f_{gui}$ actively proposes intermediate states in guided steps while $f_{syn}$ reasons freely in unguided steps -- neither is reduced to a preprocessing or translation role. 
Interleaved Granularity is realized \textit{across the schedule}: every step position $t \leq K$ is handled by transduction at some iteration and by induction at another, so no step is permanently assigned to a single paradigm. 
Search Autonomy is preserved because at iteration $j$, the inductive synthesizer reasons freely over all $K - j$ remaining steps after the last guided transition, and across the full schedule each step position is eventually approached in either reasoning mode. 
Fig.~\ref{fig:tiips_traj} illustrates this for $j=2$. 

The transductive guide $f_{\mathrm{gui}}$ is trained with teacher forcing to predict the next execution state produced by the ground-truth program at each step of the reference trajectory. 
The inductive synthesizer $f_{\mathrm{syn}}$ is trained to predict the next \ac{dsl} operation and its arguments conditioned on the current state and ground truth subtask.
Both are implemented as Transformer encoder-decoder models.
Full details on model sizes, training procedures, and domain-specific execution semantics are provided in the supplementary material.

%% file: chapters/06evaluation.tex
\section{Evaluation}
\label{sec:eval}

\paragraph{Domains}
We evaluate on three standard \ac{pbe} testbeds of increasing complexity.
RobustFill~\cite{devlin2017robustfill} is a string manipulation domain whose \ac{dsl} covers substring extraction, modification, and composition; each task has a single semantic solution trace, e.g., \texttt{Compose(ToCase(PROPER), GetFrom(' ')) | GetUpto(',')} transforms \texttt{"ITA, NAPLES"} into \texttt{"NAPLES, ITA"}.
DeepCoder~\cite{balog2016deepcoder} is a list manipulation domain with first- and higher-order functions such as \texttt{Map} and \texttt{Filter}; programs are built line by line, each applying an expression to the inputs or a prior result, and multiple semantically distinct solution traces exist per task.
LambdaBeam~\cite{shi2023exedec} extends DeepCoder with arbitrary lambda functions and conditional operations such as \texttt{If}, breaking the linear execution structure of the previous two domains and introducing branching program paths.
Together, these domains form a progression from structured, single-trace tasks to complex, multi-trace tasks with branching execution.

Following~\cite{shi2023exedec}, we evaluate on compositional generalization categories which assess the ability to generalize to novel combinations of known components~\cite{wiedemer2024compositional}. 
Five distribution shift categories are used: length generalization (increased synthesis steps), compose different concepts (novel combinations of functions from separate categories), switch concept order (reversed function sequences), compose new operations (isolated functions used in composition), and add operation functionality (new functionalities applied to known operations). 
Full domain details, train and test data generation are provided in the supplementary material.

\paragraph{Baselines} are chosen to cover each category of the taxonomy introduced in Sec.~\ref{sec:concept}, allowing us to isolate the effect of cooperation from the effects of architecture, capacity, or search diversity.
The inductive baseline uses the same setup as \ac{tiips} but without the transductive guidance model: synthesis proceeds entirely in inductive mode, with no transductive guidance.
This represents the category of execution-guided inductive search.
ExeDec~\cite{shi2023exedec} serves as the hybrid baseline that employs transduction and induction. 
It guides every step transductively: the inductive synthesizer is given a sequence of transductively prescribed target states and must realize each one, retaining only local reasoning autonomy within each subgoal. 
Thus, the only difference between \ac{tiips} and ExeDec is the interaction schedule -- in ExeDec, $f_{\mathrm{syn}}$ translates a trajectory it did not choose and cannot revise. In \ac{tiips}, it uses each transductive prediction merely as a starting point and retains full search autonomy over the remaining steps.
We do not include the BARC~\cite{li2024combining} ensemble as a separate baseline but through post-hoc analysis: the set of tasks solved by either the inductive baseline or ExeDec corresponds directly to the ensemble.
Critically, all approaches share identical model architectures and weights: $f_{\mathrm{syn}}$ is the same across all three, and $f_{\mathrm{gui}}$ is shared between \ac{tiips} and ExeDec. 
The only variable across conditions is the interaction paradigm.
We do not include a purely transductive baseline because our \ac{loo} evaluation requires executing a program on a held-out input.
Purely transductive approaches produce no such artifact and therefore cannot be evaluated under this protocol.

\paragraph{Evaluation Protocol}
All experiments are run over 5 random seeds on 1000 test tasks per seed and category, using a beam size of 10.
Results are reported as mean and standard deviation; statistical significance is assessed via paired t-tests at the 5\% level, applied per category.
Programs are evaluated under a \ac{loo} protocol: $n-1$ \ac{io} pairs are used to synthesize a program, and the held-out $n$-th pair is used to verify correctness. 
A task is solved if the program correctly maps all $n$ inputs to their respective outputs.
The step limits $J = K$ are set to 10 for the list domains and 20 for the string domain. 
More details and formal definitions can be found in the supplementary material.
Because \ac{tiips} runs $J$ iterations of the intervention schedule, each baseline receives $J$ independent attempts per task, ensuring all approaches evaluate the same total number of programs. 
Furthermore, \ac{tiips} and ExeDec can make identical total model calls per task. 
These design choices ensure that any observed performance difference reflects the interaction paradigm rather than a compute or search diversity advantage.

%% file: chapters/07results.tex
\section{Results \& Discussion}
\label{sec:results}
Our work is driven by the hypothesis that the complementarity of transduction and induction also holds on the subtask level. 
If this holds, \ac{tiips} -- as the proof-of-concept implementation of our cooperative reasoning framework -- should solve more tasks than both single-paradigm and hybrid subordination approaches.
Across all conditions, the only variable is the interaction paradigm. 

\subsection{Cooperation Improves Performance}
\paragraph{Performance Gains.}
\begin{table}[!b]
\centering
\footnotesize
\begin{tabular}{c|c|ccc}
\hline
Domain & Category & Ind. Baseline & ExeDec & \ac{tiips} \\
\hline

\multirow{7}{*}{RobustFill}
 & Test on training distribution      & 11.6 $\pm$ 0.31 & 94.8 $\pm$ 0.34 & \textbf{94.9 $\pm$ 0.36} \\
 & Length generalization              & 0.00 $\pm$ 0.00   & \textbf{91.4 $\pm$ 0.44} & \textbf{91.4 $\pm$ 0.46} \\
 & Compose different concepts         & 4.92 $\pm$ 0.77 & \textbf{98.1 $\pm$ 0.52} & \textbf{98.1 $\pm$ 0.52} \\
 & Switch concept order               & 7.96 $\pm$ 1.46 & \textbf{96.7 $\pm$ 0.92} & \textbf{96.7 $\pm$ 0.94} \\
 & Compose new operation              & 2.64 $\pm$ 0.25 & 87.1 $\pm$ 1.30 & \textbf{87.6 $\pm$ 1.24} \\
 & Add operation functionality        & 9.78 $\pm$ 0.16 & \textbf{58.3 $\pm$ 0.49} & \textbf{58.3 $\pm$ 0.53} \\
 & Generalization average             & 6.33 $\pm$ 2.92 & 86.3 $\pm$ 14.86 & \textbf{86.4 $\pm$ 14.86} \\
\hline

\multirow{7}{*}{DeepCoder}
 & Test on training distribution      & 30.4 $\pm$ 1.69 & 45.2 $\pm$ 1.70  & \textbf{47.5 $\pm$ 0.75} \\
 & Length generalization              & 0.16 $\pm$ 0.21 & 4.40 $\pm$ 0.83 & \textbf{6.34 $\pm$ 1.60} \\
 & Compose different concepts         & 15.2 $\pm$ 0.81 & 19.1 $\pm$ 1.46 & \textbf{26.0 $\pm$ 1.59} \\
 & Switch concept order               & 16.6 $\pm$ 1.23  & 14.3 $\pm$ 4.17 & \textbf{26.6 $\pm$ 1.78} \\
 & Compose new operation              & 17.1 $\pm$ 1.93  & 11.6 $\pm$ 1.62 & \textbf{24.3 $\pm$ 2.53} \\
 & Add operation functionality       & 11.0 $\pm$ 1.09  & 9.76 $\pm$ 1.16 & \textbf{17.0 $\pm$ 0.70} \\
 & Generalization average             & 12.0 $\pm$ 6.52  & 11.8 $\pm$ 5.36 & \textbf{20.1 $\pm$ 7.98} \\
\hline

\multirow{7}{*}{LambdaBeam}
 & Test on training distribution      & 42.74 $\pm$ 2.81 
 & 50.95 $\pm$ 1.50 & \textbf{57.60 $\pm$ 0.91} \\
 & Length generalization              & 7.45 $\pm$ 1.34 
 & 13.74 $\pm$ 2.04 & \textbf{16.58 $\pm$ 1.73} \\
 & Compose different concepts         & 21.47 $\pm$ 2.56 
 & 33.00 $\pm$ 3.08 & \textbf{39.27 $\pm$ 3.66} \\
 & Switch concept order               & 19.26 $\pm$ 1.37 
 & 17.76 $\pm$ 1.32 & \textbf{26.81 $\pm$ 0.87} \\
 & Compose new operation               & 18.94 $\pm$ 2.81 
 & 31.10 $\pm$ 1.50 & \textbf{35.01 $\pm$ 1.19} \\
 & Add operation functionality       & 16.71 $\pm$ 3.54 
 & 20.98 $\pm$ 3.32 & \textbf{31.09 $\pm$ 3.87} \\
 & Generalization average             & 16.77 $\pm$ 5.48 
 & 23.31 $\pm$ 7.98 & \textbf{29.75 $\pm$ 8.28} \\
\hline

\end{tabular}
\caption{Mean accuracy $\pm$ std. dev. across domains and categories. Bold indicates the best performing result. \ac{tiips} significantly outperforms ExeDec in the DeepCoder and LambdaBeam domain and performs on par with ExeDec on the RobustFill domain.}
\label{tab:acc}
\end{table}
Table~\ref{tab:acc} reports task accuracy across all three domains and out-of-distribution categories. 
\ac{tiips} consistently outperforms both baselines in DeepCoder and LambdaBeam, with all improvements being statistically significant ($p < 0.05$; full statistics in supplementary material).
In DeepCoder, \ac{tiips} improves over ExeDec on average by approximately 10 percentage points and over the Ind. Baseline by approximately 20 percentage points. 
In LambdaBeam, the corresponding improvements are 7 and 13 percentage points. 
Notably, in several categories of both domains, e.g., Switch Concept Order, the Ind. Baseline matches or outperforms ExeDec.
This highlights the subordination liability: in complex, multi-trace domains where the transductive guide is frequently off-target, forcing the synthesizer to follow a wrong intermediate prediction prunes the correct solution from the search space before induction begins. 
An unconstrained inductive model, retaining full search autonomy, avoids this failure entirely. 
In LambdaBeam, this effect is further demonstrated, yet to a lesser extent.
In RobustFill, \ac{tiips} and ExeDec perform on par (mean difference $+0.1$ percentage points, not statistically significant across categories). 

\begin{figure}[!b]
    \centering
    \begin{minipage}{0.32\linewidth}
        \centering
        \includegraphics[width=\linewidth]{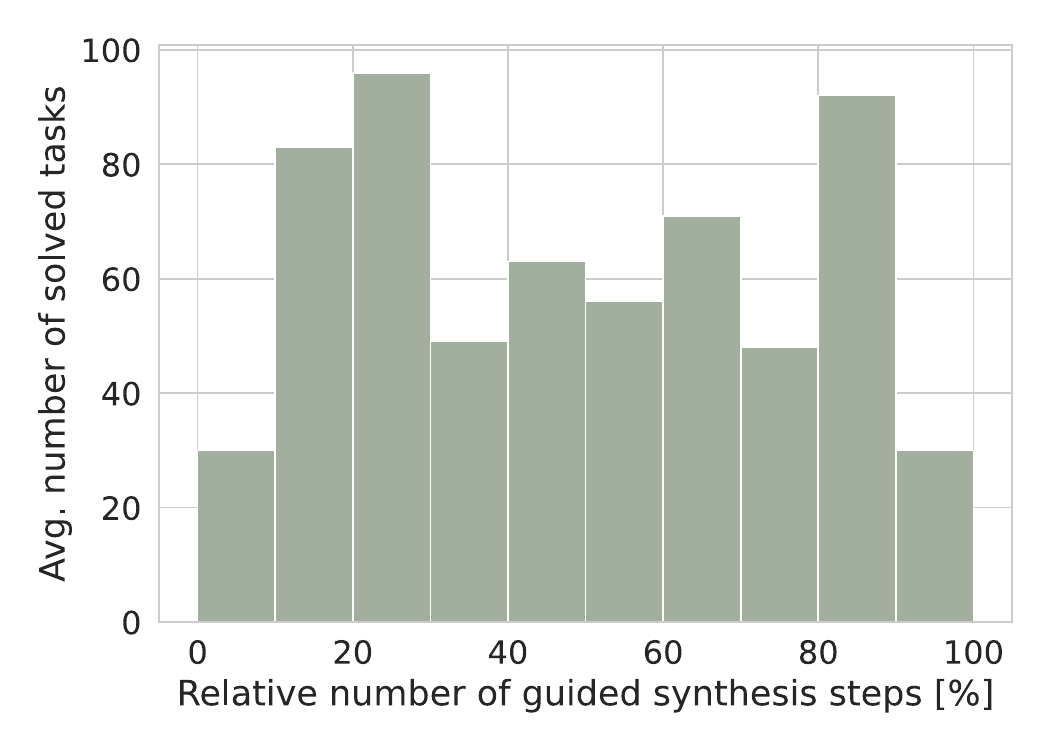}
        \\ \footnotesize (a) DeepCoder
    \end{minipage}
    \begin{minipage}{0.32\linewidth}
        \centering
        \includegraphics[width=\linewidth]{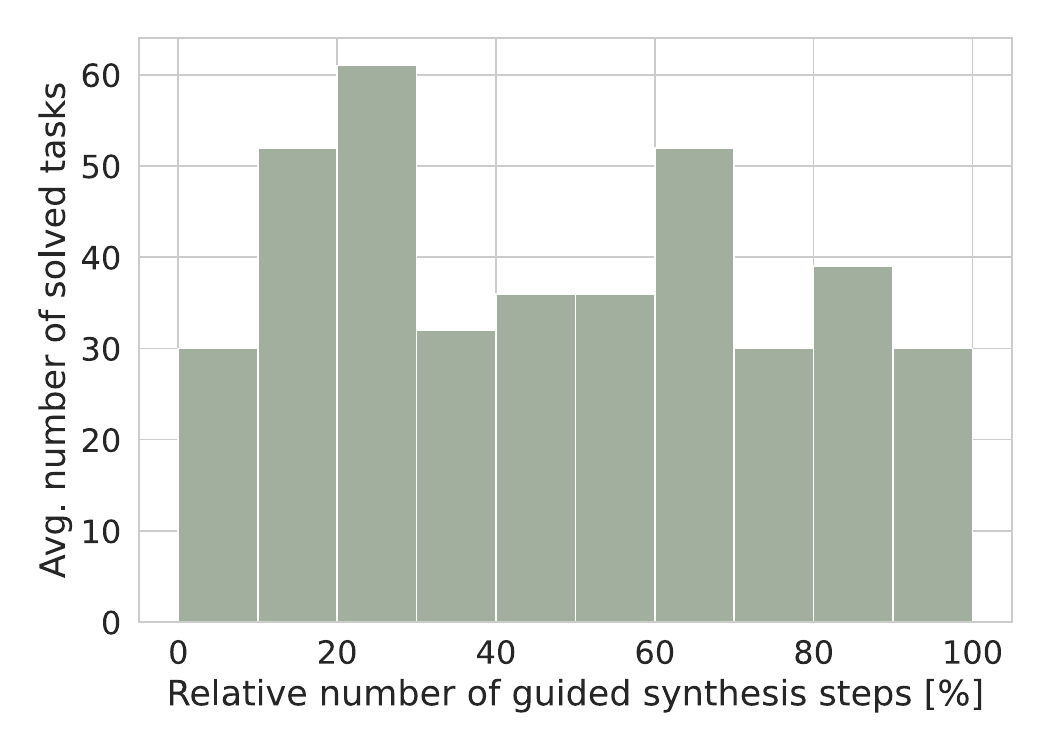}
        \\ \footnotesize (b) LambdaBeam
    \end{minipage}
    \begin{minipage}{0.32\linewidth}
        \centering
        \includegraphics[width=\linewidth]{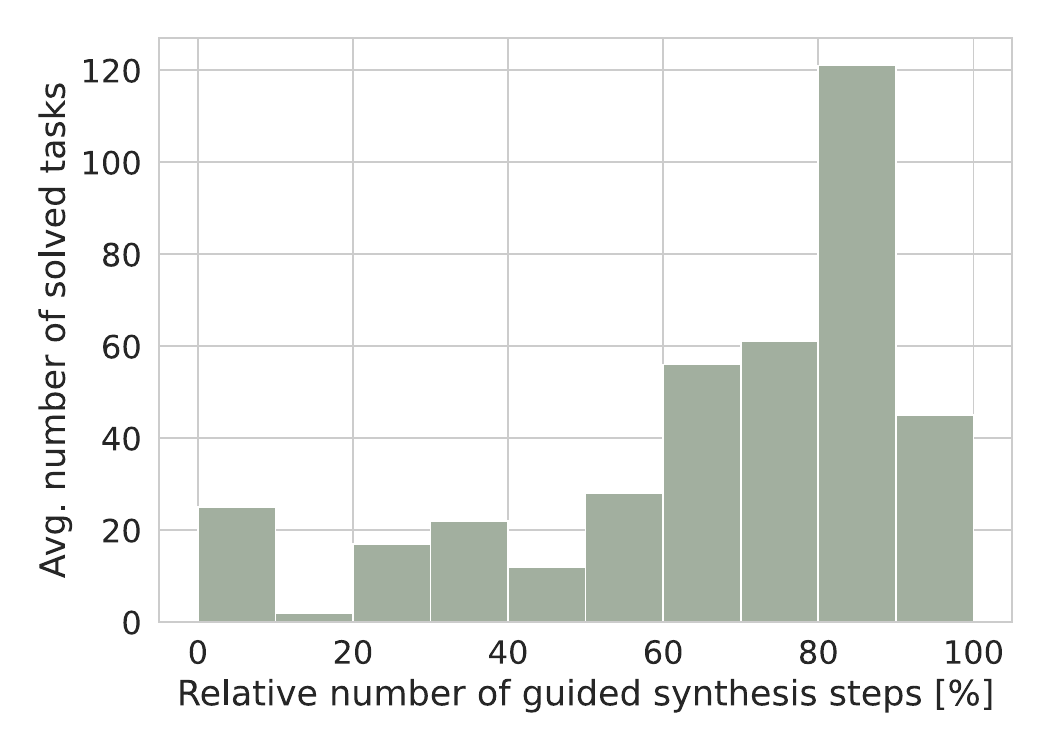}
        \\ \footnotesize (c) RobustFill
    \end{minipage}
    \caption{Distribution of transductive intervention levels $j$ at which tasks are solved by \ac{tiips}, aggregated across all categories and seeds. In the list domains gains are spread uniformly across the full trajectory depth, confirming syntactic bottlenecks occur throughout synthesis. In RobustFill the distribution is right-skewed, reflecting natural convergence to full guidance in a single-trace domain.}
    \label{fig:schedule}
\end{figure}

\paragraph{When Does Cooperation Activate?}
This convergence is not surprising -- it is in fact what the incremental intervention schedule predicts. 
In a single-trace domain, the transductive guide reliably sketches an accurate decomposition and the synthesizer rarely needs to exercise its search autonomy, making cooperation and subordination equivalent.
Fig.~\ref{fig:schedule} confirms this directly: in RobustFill, the distribution is strongly right-skewed, with most tasks solved at or near $j=J$.
The schedule naturally escalates to full guidance for most tasks, recovering ExeDec-like behavior, while in DeepCoder and LambdaBeam the cooperative regime $0 < j < J$ remains active across the full trajectory depth. 
Notably, a non-trivial fraction of tasks are solved at $j=0$ in the list domains, reflecting that transductive guidance is not always beneficial or necessary -- the schedule attempts induction first and escalates only when needed.
Since $j=0$ recovers the Ind. Baseline and $j=J$ recovers ExeDec, all performance gains are attributable exclusively to the cooperative regime $0 < j < J$, and given the compute alignment of the approaches, any observed difference reflects the interaction paradigm alone.

\paragraph{Cooperation Does the Trick.}
\label{sec:results_venn}
\begin{figure}[!b]
    \centering
    \subcaptionbox{DeepCoder\label{fig:venn_deepcoder}}{
        \includegraphics[width=0.47\linewidth]{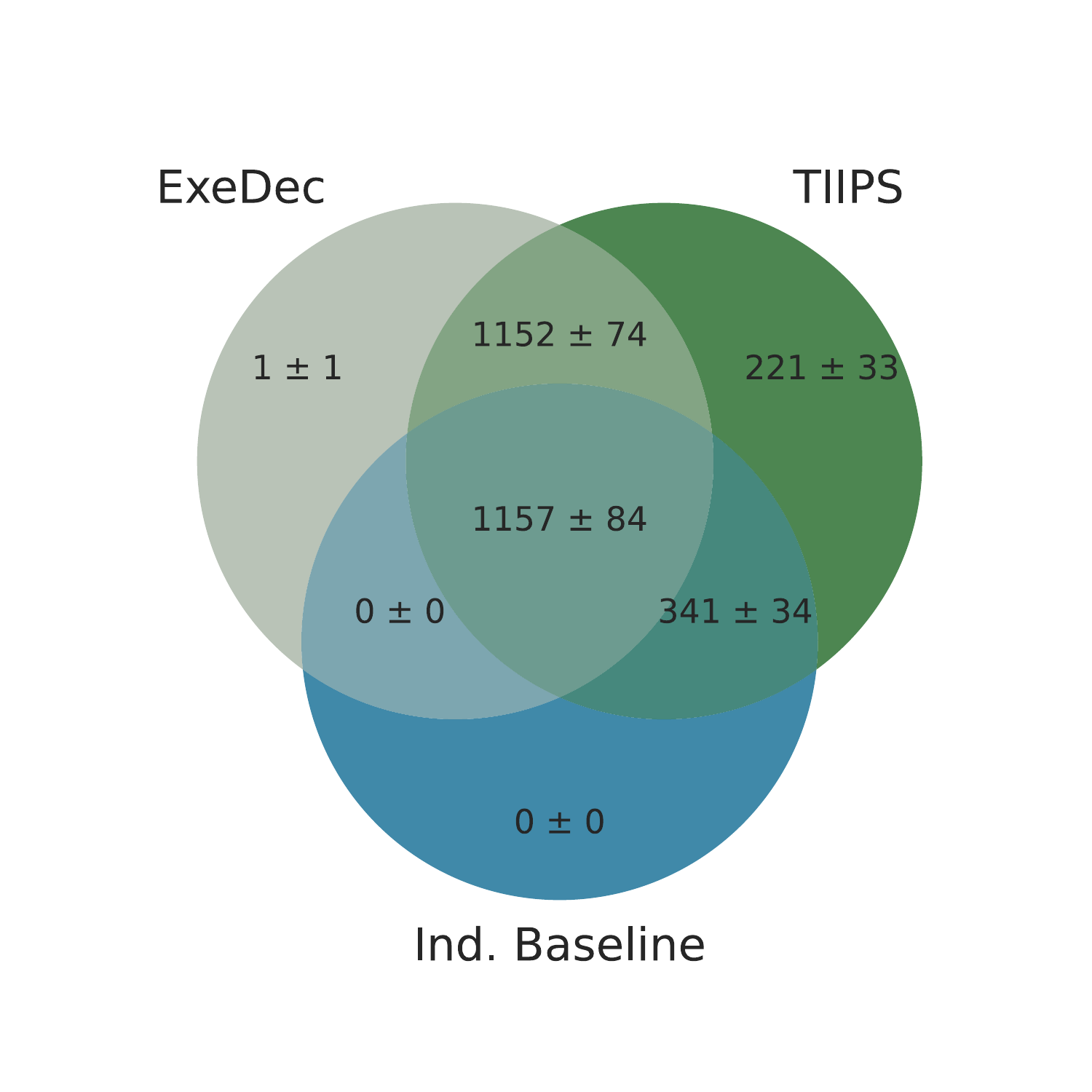}
    }
    \subcaptionbox{LambdaBeam\label{fig:venn_lambdabeam}}{
        \includegraphics[width=0.47\linewidth]{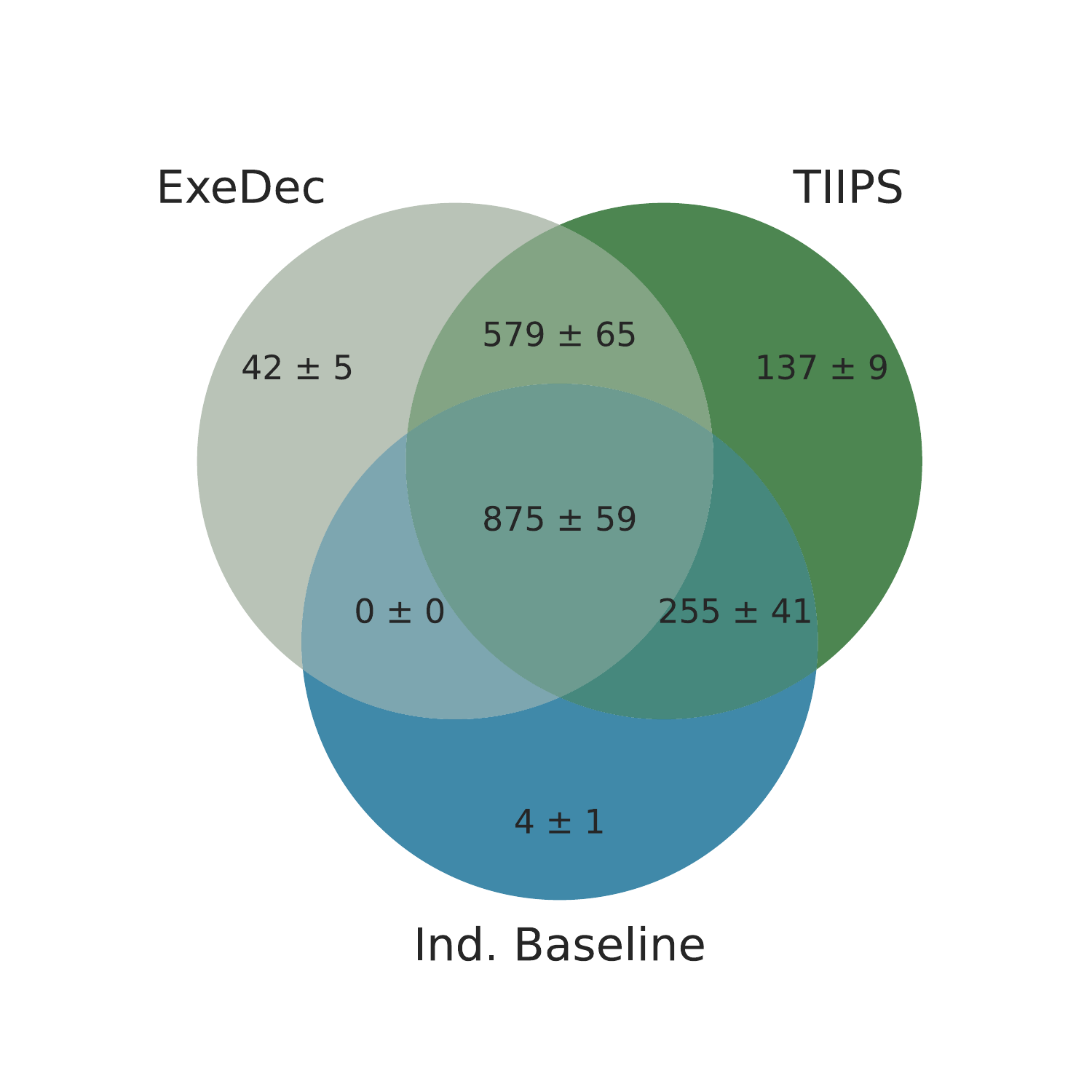}
    }
    \subcaptionbox{RobustFill\label{fig:venn_robustfill}}{
        \includegraphics[width=0.47\linewidth]{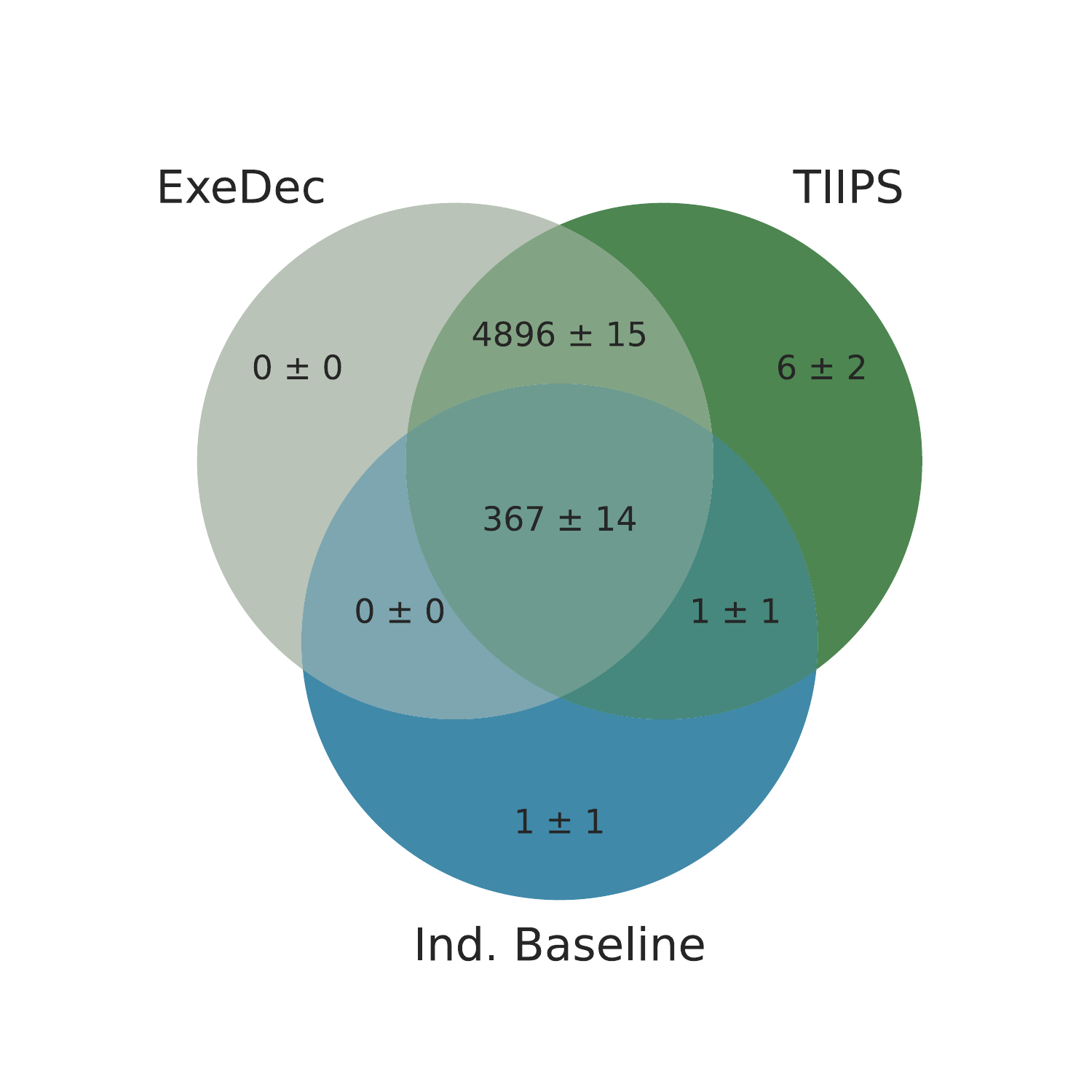}
    }
    \caption{Venn diagrams showing overlap of solved tasks across all categories and seeds between \ac{tiips}, ExeDec, and the Ind. Baseline. \ac{tiips} solves a large amount of tasks that are not solved by either of the two baselines (union of the baselines = Ensemble) -- demonstrating that cooperative interaction does the trick -- not just ensembling. Unlike the list domains, \ac{tiips} and ExeDec converge almost entirely in the RobustFill domain, consistent with the cooperative framework's prediction that cooperation and subordination become equivalent when the transductive guide is reliable.}
    \label{fig:venn}
\end{figure}
Figure~\ref{fig:venn} shows Venn diagrams of tasks solved by each approach, aggregated across all categories and seeds. 
Three findings stand out, each speaking to a distinct aspect of the cooperative framework.
First, the Ind. Baseline and ExeDec solve largely distinct task sets, directly confirming the subtask-level complementarity illustrated in Fig.~\ref{fig:complementarity}. 
In DeepCoder, ExeDec solves approximately 1,150 tasks the Ind. Baseline does not, while the baseline solves approximately 340 tasks ExeDec does not. 
In LambdaBeam, the corresponding figures are 600 and 260.
Each paradigm has genuine strengths the other lacks -- neither dominates globally.
Second, \ac{tiips} covers the vast majority of tasks solved by either baseline, demonstrating that cooperation does not sacrifice the individual strengths of either paradigm. 
The cooperative solver subsumes both.
Third and most critically \ac{tiips} solves a substantial number of tasks that are unreachable by either baseline individually or in combination. 
In DeepCoder, \ac{tiips} exclusively solves $221 \pm 33$ tasks (approximately 8\% of all solved tasks), while the union of both baselines -- the ensemble ceiling -- exclusively solves only $1 \pm 1$.
In LambdaBeam, the figures are $137 \pm 9$ and $46 \pm 6$ respectively.
In RobustFill, consistent with the predicted convergence, the exclusive gain shrinks to $6 \pm 2$ tasks. 
The pattern differs markedly from the list domains: ExeDec solves approximately 4,900 tasks the Ind. Baseline does not, while the baseline solves only a single task ExeDec does not -- confirming that in a single-trace domain subordination rarely causes harm, as the transductive guide is sufficiently reliable. 
\ac{tiips} covers virtually all tasks solved by ExeDec and adds $6 \pm 2$ exclusive tasks.
This small but non-zero number indicates that even in a near-optimal guidance regime, occasional transductive mispredictions occur and \ac{tiips}'s preserved search autonomy allows recovery where ExeDec cannot -- precisely what the cooperative framework predicts.
We call this the \emph{cooperative dividend}: the set of tasks that are unreachable by the non-cooperative approaches.
The cooperative dividend rules out two alternative explanations: It can not be explained by \ac{tiips} functioning as a variable-length prefix of ExeDec.
If that were the case, \ac{tiips} would be a strict subset of ExeDec's solved tasks. 
Nor can it be explained by \ac{tiips} functioning as an ensemble: it solves tasks that neither baseline solves individually nor in combination, which by definition lies beyond the reach of any ensemble of the two. 
Instead, \ac{tiips} succeeds on these tasks because after each guided step the synthesizer retains the freedom to diverge from the transductively prescribed trajectory and explore the full program space -- a structural capability that no ensemble of isolated solvers can replicate, directly validating the search autonomy preservation criterion.

\subsection{Cooperation improves Robustness}
\begin{figure}[t]
    \centering
    \begin{minipage}{0.45\textwidth}
        \centering
        \includegraphics[width=\linewidth]{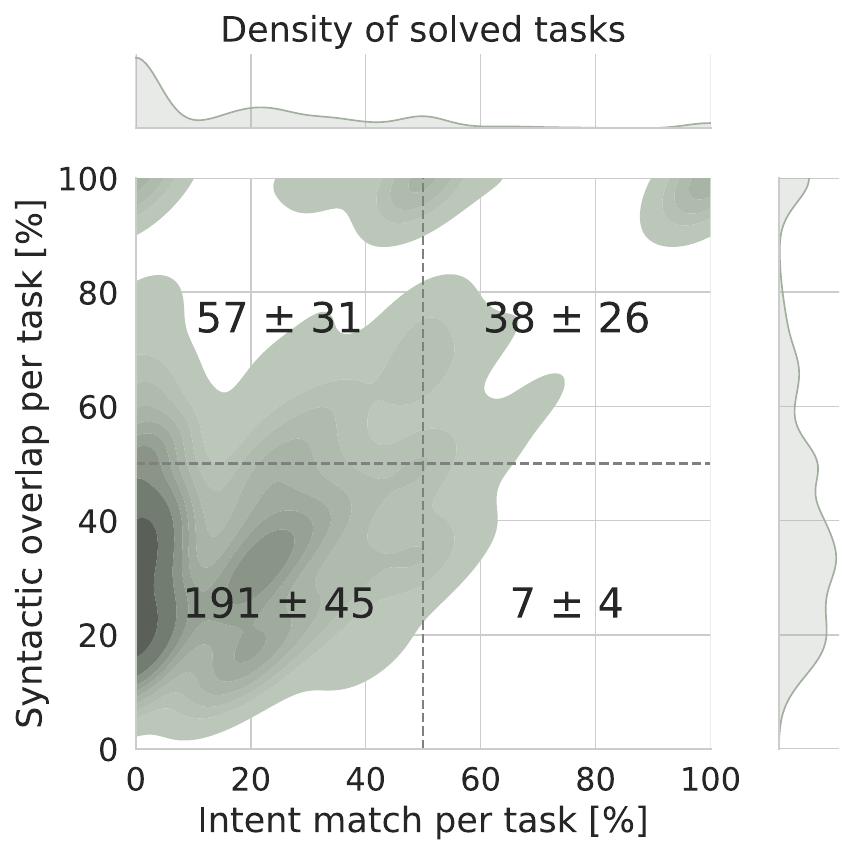}
        \\ \footnotesize (a) ExeDec
    \end{minipage}
    \hfill
    \begin{minipage}{0.45\textwidth}
        \centering
        \includegraphics[width=\linewidth]{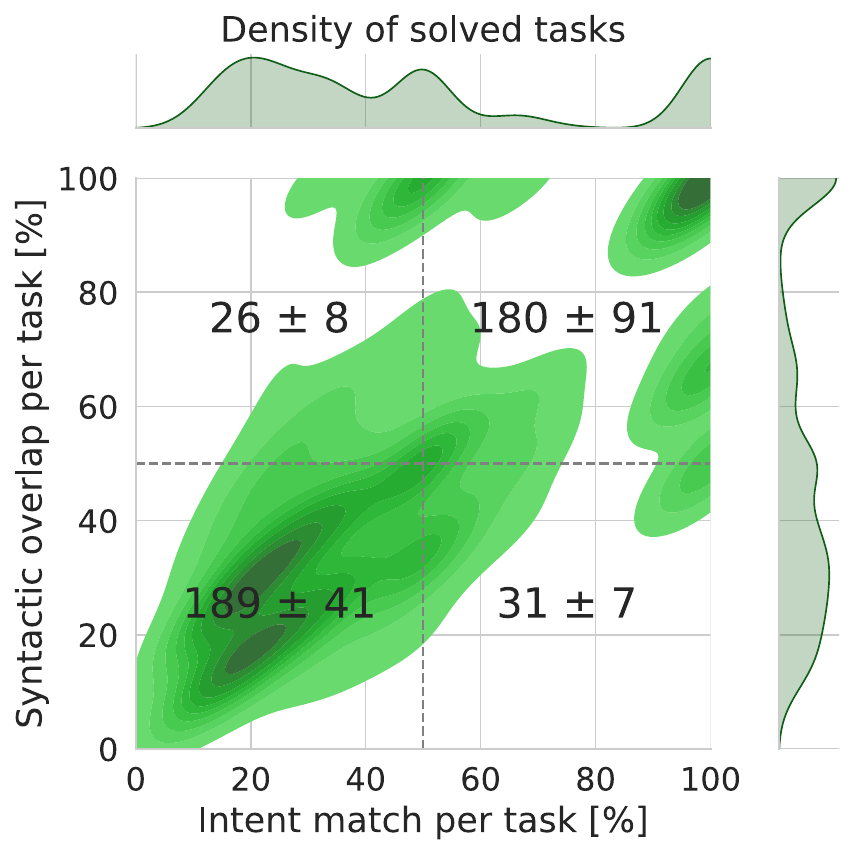}
        \\ \footnotesize (b) \ac{tiips}
    \end{minipage}
    \caption{Density plot of tasks solved by \ac{tiips} and ExeDec in the DeepCoder domain, grouped by intent match and syntactical overlap. \ac{tiips} solutions show higher intent and syntactical overlap with the ground truth. Intent match measures the overlap between predicted/executed and ground-truth subtask outputs, while syntactical overlap reflects the structural similarity of predicted programs to ground-truth. Displayed numbers show the average number of task solutions and their standard deviation.}
    \label{fig:density_dc}
\end{figure}

Figure~\ref{fig:density_dc} shows kernel density plots of tasks solved by \ac{tiips} and ExeDec in the DeepCoder domain, distributed by intent match and syntactic overlap with the ground-truth solution.
A solution in the top-right quadrant captures the intended logic and syntax rather than an incidental pattern, and is therefore more likely to reproduce the desired program behavior on unseen samples.

\ac{tiips} solutions concentrate strongly in the top-right quadrant ($180 \pm 91$ tasks vs. $38 \pm 26$ for ExeDec) and exhibit a pronounced right-skew on the intent match axis, indicating that \ac{tiips} more frequently recovers the correct intermediate semantic states along the solution trajectory.
ExeDec solutions, by contrast, distribute broadly across the lower quadrants, with a notable cluster in the top-left ($57 \pm 31$) -- programs that match the ground-truth syntactically but miss the intended semantic trace. 
This pattern is the fingerprint of the subordination failure mode: when the transductive guide predicts a slightly off-target intermediate state, ExeDec's synthesizer must generate convoluted programs to satisfy that wrong constraint, producing code that passes the training examples but might deviate from the canonical transformation. 
The result is low semantic grounding and brittle generalization to unseen test samples.
\ac{tiips} avoids this failure through search autonomy. 
Because the synthesizer is not locked into the transductive guide's prescription after each step, it can effectively preempt a misleading prediction: rather than generating a program that reaches a wrong intermediate state exactly, it finds a \ac{dsl}-natural subprogram that arrives at a more semantically grounded state than the guide originally proposed.
\ac{tiips} does not merely find a solution that satisfies the \ac{io} examples -- it more frequently finds the \emph{intended} solution. 
This indicates that a cooperative solver is not simply more powerful than its components in isolation or ensembling but it is more reliably correct.

%% file: chapters/08conclusion.tex
\section{Conclusion}
We introduced cooperative transductive-inductive problem solving, a framework in which inductive and transductive reasoning interleave dynamically, with neither paradigm entirely controlling the solution trajectory. 
The implications of cooperative reasoning extend well beyond the \ac{pbe} setting.
A prominent trend in recent research has been to move away from explicit program generation on the grounds that programs are too difficult to find~\cite{chollet2024arc}. 
\ac{tiips} demonstrates that the reasoning abilities of symbolic solvers need not be sacrificed to obtain the pattern-matching power of transductive inference: the two can cooperate. 
Rather than abandoning programs, the field should explore how symbolic and neural reasoning can be made to cooperate.

\paragraph{Limitations} Our evaluation is conducted within the program synthesis setting, where the terminal objective is a symbolic program artifact.
While we argue that the cooperative principle is paradigm-agnostic, its applicability to reasoning tasks that do not require a program as final output remains to be demonstrated empirically.
At instance level, \ac{tiips} currently has no mechanism to detect when the inductive synthesizer has stalled and guidance would be beneficial.
Replacing this static schedule with an adaptive mechanism is a natural direction for future work: such a mechanism could condition on signals from the inductive search process -- such as beam diversity, search depth, or solution consistency across examples -- to decide at each trajectory step whether to continue inductively or trigger a transductive reset.
Taken further, this could be cast as a learned routing policy moving closer to the full cooperative ideal described by the framework. 
These are structural limitations of the schedule-based instantiation, not of the cooperative framework itself.
Indeed, the fact that even this minimal schedule with small Transformer models yields consistent and substantial gains speaks to the robustness of the underlying concept rather than to \ac{tiips} specifically.

\section*{Acknowledgements}
This research was supported in part by the German Federal Ministry for Economic Affairs and Energy (BMWE).